\documentstyle[multicol,prb,aps]{revtex}
\begin{document}
\draft
\preprint{HEP/123-qed}
\title{THEORY OF SPIN INJECTION}
\author{EMMANUEL I. RASHBA\\} 
\address{Department of Physics, MIT, Cambridge, Massachusetts 02139, and 
\\Department of Physics, The State University of New York at Buffalo, Buffalo, NY 14260\\}  

\maketitle
\begin{center}
{\it Email: erashba@mailaps.org}
\end{center}
\begin{abstract}
Diffusive theory of spin injection is reviewed and a number of new results is presented for the dc and ac regimes. They were derived by means of the $\gamma$-technique allowing to simplify the calculations by choosing the spin injection coefficients through different interfaces as the basic variables. The prospects for increasing spin injection by using resistive spin-selective contacts are emphasized and spin non-conserving contacts are introduced. Finding the basic parameters of a junction from the ac data is discussed.
\end{abstract}
\begin{multicols}{2}
\narrowtext

\subsection*{1~~Introduction}             

Spin injection is believed to be the key to many new phenomena and applications in the field of the spin-polarized electron transport.\cite{Wolf,DS} Ferromagnetic metals are robust sources of spin-polarized electrons applicable in a wide range of temperatures and requiring no external magnetic field, while semiconductor microstructures are well suited for operating the transport of the injected spin-polarized electrons. However, the first experimental studies failed to achieve a considerable spin polarization degree of the current injected from metallic ferromagnets into semiconductors, and the concept of the ``conductivity mismatch" provided a natural explanation of that failure. I argue in what follows that for a properly designed ferromagnet-semiconductor junction the difference in the conductivities of the different elements of the junction becomes an advantage rather than an obstacle for efficient spin injection. For this purpose {\it resistive spin-selective contacts} should be employed.

Spin injection from a ferromagnetic (F) source into a semiconductor (more generally, into a normal conductor, N) across a resistive tunnel or Schottky contact (T) is controlled by three competing resistances: $r_F$ and $r_N$, the effective resistances of F and N conductors, and $r_c$, a contact resistance. The spin injection coefficient $\gamma$ of the junction is controlled by the largest of these three resistances. The resistance $r_c$ is very small for a ``perfect" contact, $r_c\approx 0$, and $r_F\ll r_N$ when N is a semiconductor and F is a metal. Under these conditions, the spin non-polarized semiconductor controls the injection and $\gamma\sim r_F/r_N\ll 1$, hence, {\it perfect contacts are ill fit for the role of spin emitters.} However, experimental data on the spin injection from magnetic STM tips and similar sources show convincingly that the contact resistance is strongly spin dependent.\cite{Al95} {\it When} $r_c\agt r_N, r_F$, the contact resistance gains control over the spin injection across the junction, and {\it the spin selectivity of the contact becomes the major factor controlling $\gamma$}.\cite{R00} Recent reports on a dramatic increase in $\gamma$ by using resistive contacts have confirmed this concept.\cite{exp}

\subsection*{2~~Diffusive theory: FTN-junction}              

The theory of spin injection takes a rather different form depending on the transport mechanism across the semiconductor (diffusive, ballistic, etc.). The diffusive approach is the basic toy model of the theory because (i) it provides a general insight on the problem, (ii) is formulated in terms of the basic physical parameters, and (iii) is most simple and results in explicit analytical formulae. The conclusions listed in the Introduction and the discussion that follows below are based on that approach. Some of the results may still remain valid even when the criteria of the diffusive approach are not fulfilled.\cite{viol} Special advantages of the ballistic regime and perfect contacts that are anticipated\cite{Kir01} are outside of scope the this paper.

The basic variables of the diffusion theory are the electrochemical potentials, $\zeta_{\uparrow,\downarrow}(x)$, and the currents, $j_{\uparrow,\downarrow}(x)$, of up- and down-spin electrons, respectively. They obey the standard equations $j_{\uparrow,\downarrow}(x)=\sigma_{\uparrow,\downarrow}\nabla\zeta_{\uparrow,\downarrow}(x)$ where $\sigma_{\uparrow,\downarrow}$ are the conductivities and the continuity equations for the currents $j_{\uparrow,\downarrow}(x)$ that include  spin relaxation times $\tau_s$. All these quantities should also bear the indeces F or N for the F- and N-regions. More attention should be paid to the boundary conditions. When a contact, at the point $x=0$, is spin conserving as is usually supposed then $j^F_{\uparrow,\downarrow}(0)=j^N_{\uparrow,\downarrow}(0)\equiv j_{\uparrow,\downarrow}$. For resistive contacts needed to achieve an efficient spin injection the potentials $\zeta_{\uparrow,\downarrow}(x)$ are discontinuous at $x=0$ and related to the currents as
\begin{equation}
j_{\uparrow,\downarrow}=\Sigma_{\uparrow,\downarrow}(\zeta^N_{\uparrow,\downarrow}(0)-\zeta^F_{\uparrow,\downarrow}(0)),
\label{eq1}
\end{equation}
where $\Sigma_{\uparrow,\downarrow}$ are contact conductivities for up- and down-spins. That is Eq.~(\ref{eq1}) that makes a critical difference between the spin injection and the Shockley's theory of $p-n$-junctions where $\zeta$'s are continuous.

Solving these equations for an isolated FTN-junction with unlimited F- and N-regions is straightforward. For the spin infection coefficient
\begin{equation}
\gamma=(j_\uparrow-j_\downarrow)/J,~~J=j_\uparrow+j_\downarrow,
\label{eq2}
\end{equation}
it provides a simple result\cite{R00,HZ97}
\begin{equation}
\gamma=[r_c(\Delta\Sigma/\Sigma)+r_F(\Delta\sigma/\sigma_F)]/r_{FN}~.
\label{eq3}
\end{equation}
Here $\Sigma=\Sigma_\uparrow+\Sigma_\downarrow$, $\Delta\Sigma=\Sigma_\uparrow-\Sigma_\downarrow$ and $\sigma_F=\sigma_\uparrow+\sigma_\downarrow$, $\Delta\sigma=\sigma_\uparrow+\sigma_\downarrow$ describe the total conductivities and spin selectivities of the contact and the ferromagnet, respectively. The denominator 
$r_{FN}=r_F+r_c+r_N$ is a sum of three effective resistances $r_F=\sigma_F L_F/4\sigma_\uparrow\sigma_\downarrow$, $r_N=L_N/\sigma_N$, and $r_c=\Sigma/{4\Sigma_\uparrow\Sigma_\downarrow}$ of the ferromagnet, the normal conductor, and the contact, respectively. $L_F$ and $L_N$ are the spin diffusion lengths in F- and N-regions while $\sigma_N$ is the N-region conductivity. For a perfect contact, $r_c=0$, this result has been known for long.\cite{vS87}

An important conclusion follows from Eq.~(\ref{eq3}). If $r_c=0$ then $\gamma\sim r_F/r_N$, hence, $\gamma\ll 1$ whenever $r_F/r_N\ll 1$. Because this is the case for a contact of a ferromagnetic metal and a semiconductor, the conductivity mismatch concept\cite{Sch00} and a pessimistic prognosis for such spin sources follow immediately. However, if the contact is both resistive, $r_c\agt r_N, r_F$, and spin selective, $\Delta\Sigma\sim\Sigma$, then $\gamma$ is about $\gamma\sim\Delta\Sigma/\Sigma$ and this ratio can be rather high. Therefore, {\it the contacts that are both resistive and spin selective can remedy the problem}. The restriction imposed on $r_c$ by this criterion is rather mild: $r_c$ should only exceed the resistances $r_F$ and $r_N$ that are inherent in the system. Under these conditions {\it the spin selectivity of the contact} $\Delta\Sigma/\Sigma$ rather than $\Delta\sigma/\sigma_F$ {\it becomes the critical factor controlling the spin injection}.

When a resistive contact has internal magnetic degrees of freedom, then its spin selectivity can be controlled by the spin polarized current injected from a ferromagnetic electrode. This idea has been elaborated in Ref.~\onlinecite{ion} for a magnetic ion doped quantum dot (microcrystal) connected to the F and N leads.

Similar but more elaborate calculations result in the resistance of a F-N-junction $R=\Sigma^{-1}+R_{\rm n-eq}$ where 
\end{multicols}
\widetext 
\begin{equation}
 R_{\rm n-eq}={1\over{r_{FN}}}\{ r_N\left[r_c(\Delta\Sigma/\Sigma)^2
+r_F(\Delta\sigma/\sigma_F)^2 \right]
+r_c r_F\left[(\Delta\Sigma/\Sigma)-(\Delta\sigma/\sigma_F)\right]^2 \}. 
\label{eq4}
\end{equation}
\begin{multicols}{2}
\noindent
$\Sigma^{-1}$ is the equilibrium part of the resistance while $R_{\rm n-eq}$ is its nonequilibrium part that vanishes when both $L_F, L_N\rightarrow 0$. Remarkably, the right hand side of Eq.~(\ref{eq4}) is evidently positive. Therefore, {\it spin injection enlarges the resistance of a junction}. This property is rather general.

\subsection*{3~~$\gamma$-technique}              

Solving the equations for an isolated spin-conserving FTN-junction is a relatively simple problem and there is a complete agreement between the results reported for $\gamma$ by different authors. However, when it comes to more involved systems including, e.g., a junction with two FTN-contacts, or to spin non-conserving junctions, the number of equations increases and the calculations get highly cumbersome. Apparently, it is why there exist controversies in the results derived for FNF- and FTNTF-junctions by different authors, and the problem of spin non-conserving junctions has not been approached until now. I emphasize that the equations for these systems are still elementary, hence, the problem is completely technical. For this purpose I have developed a special technique ($\gamma$-technique) that allows to organize calculations in such a way that they get simpler and, therefore, the results become more reliable.  

In the $\gamma$-technique (i) the coefficients of spin injection, $\Gamma$'s, through the different interfaces (or through the left and right boundaries of the same spin non-conserving contact) become the basic variables, (ii) the external parts of the junction are eliminated and their parameters are absorbed into the boundary values of $\zeta$'s at the interfaces, (iii) these $\zeta$'s are expressed through $\Gamma$'s, and (iv) the self-consistency condition for $\Gamma$'s is derived. Of course, $\Gamma$'s for a system with a finite N-region differ from $\gamma$'s found for an unlimited FTN-junction, Eq.~(\ref{eq3}), but they can be expressed in terms of those $\gamma$'s. These equations are concise when written in appropriate notations. The junction resistance $R$ and the spin valve effect $\Delta R$ can be expressed in the same terms. All results presented below were derived by this procedure.

\subsection*{4~~FTNTF-junction}              

When both FTN-contacts are spin conserving, the system of equations for all $\zeta(x)$'s and $j(x)$'s in F- and N-regions, including the boundary conditions, can be split into two systems. The first system includes only differences $\zeta_\uparrow(x)-\zeta_\downarrow(x)$ and $j_\uparrow(x)-j_\downarrow(x)$ that can be related to $\Gamma_L$ and $\Gamma_R$, the spin injection coefficients through the left and right contact, respectively. The $\gamma$-technique results in the following equations for $\Gamma_L$ and $\Gamma_R$:
\begin{eqnarray}
r_{FN}^L(d)\Gamma_L-\left\{r_N/\sinh(d/L_N)\right\}\Gamma_R&=&r_{FN}^L\gamma_L,\nonumber \\
-\left\{r_N/\sinh(d/L_N)\right\}\Gamma_L+r_{FN}^R(d)\Gamma_R&=&r_{FN}^R\gamma_R,
\label{eq5}
\end{eqnarray}
where $d$ is the width of the N-region and
\begin{equation}
r_{FN}^{L(R)}(d)=r_F^{L(R)}+r_c^{L(R)}+r_N\coth(d/L_N).
\label{eq6}
\end{equation}
The simple system of two equations, Eq.~(\ref{eq5}), with $\gamma_{L,R}$ of Eqs.~(\ref{eq3}) in the right hand sides, describes completely the spin injection through an asymmetric FTNTF-junction, parameters of both ferromagnets and both contacts are completely independent. As applied to a symmetric junction, various injection regimes have been discussed in Ref.~\onlinecite{Fert} in the framework of the traditional approach.

Calculating the junction resistance $R$ is a more challenging problem. For this purpose one should solve the equation for the symmetric combination of the electrochemical potentials $Z(x)=[\zeta_\uparrow(x)+\zeta_\downarrow(x)]/2$, apply Eq.~(\ref{eq5}), and take advantage of the fact that the total drops of $Z(x)$ and of the electrical potential $\varphi(x)$ across the junction are exactly equal. For a symmetric junction the resistance $R=2\Sigma^{-1}+R_{\rm n-eq}$, and its nonequilibrium part equals
\end{multicols}
\widetext 
\begin{equation}
 R_{\rm n-eq}(\gamma_L,\gamma_R)=2[r_F(\Delta\sigma/\sigma_F)^2
+r_c(\Delta\Sigma/\Sigma)^2]-2(r_{FN}^2/{\cal D})[\gamma^2r_{FN}(d)
+\gamma_L\gamma_Rr_N/\sinh(d/L_N)].
\label{eq7}
\end{equation}
\begin{multicols}{2}
\noindent
Here ${\cal D}=(r_F+r_c)^2+r_N^2+2r_N(r_F+r_c)\coth(d/L_N)$. 

Experimentally two basic configurations are of interest, with the parallel and antiparallel magnetization of the leads when $\gamma_L=\gamma_R$ or $\gamma_L=-\gamma_R$, respectively. In both cases $|\gamma_L|=|\gamma_R|\equiv\gamma $. It is a remarkable property of Eq.~(\ref{eq7}) that the valve effect $\Delta R=R(\uparrow\downarrow)-R(\uparrow\uparrow)$ [i.e., the difference in $R$ for the antiparallel and parallel configurations] comes exclusively from its very last term proportional to the mixed product $\gamma_L\gamma_R$. That is one of the reasons why the representation of $R$ in terms of $\gamma$'s is so advantageous. With Eq.~(\ref{eq3}) taken into account, the explicit expression for the spin valve effect is
\end{multicols}
\widetext 
\begin{equation}
\Delta R= {{4r_N\left(r_c{{\Delta\Sigma}\over\Sigma}
+r_F{{\Delta\sigma}\over{\sigma_F}}\right)^2}\over 
{[(r_F+r_c)^2+r_N^2]\sinh(d/L_N)+2r_N(r_F+r_c)\cosh(d/L_N)}}.
\label{eq8}
\end{equation}
Using Eqs.~(\ref{eq3}) and (\ref{eq7}), one can also find the nonequilibrium resistance $R_{\rm n-eq}(\uparrow\uparrow)$. When $r_c=0$, it equals
\begin{equation}
 R_{\rm n-eq}(\uparrow\uparrow)=2r_Fr_N\left(\Delta\sigma/\sigma_F\right)^2
{{r_N\sinh(d/L_N)+r_F[\cosh(d/L_N)-1]}\over{(r_F^2+r_N^2)\sinh(d/L_N)
+2r_Fr_N\cosh(d/L_N)}}.
\label{eq9}
\end{equation}
\begin{multicols}{2}
\noindent
Both Eqs.~(\ref{eq8}) and (\ref{eq9}) are new. To the best of my knowledge, they differ from various equations available in literature. More general equations will be published elsewhere.

It is seen from Eq.~(\ref{eq9}) that $ R_{\rm n-eq}(\uparrow\uparrow)>0$.
This property established in Ref.~\onlinecite{R00} has been observed experimentally in Ref.~\onlinecite{Sch01} by changing gradually the magnetization of semimagnetic electrodes.

\subsection*{5~~Measuring basic parameters}              

Spin injection coefficients depend critically on the relative values of a number of different parameters related to the bulk and the interfaces. These are the effective resistances ($r_F$, $r_N$, and $r_c$), the parameters on which they depend like spin diffusion lengths ($L_F$ and $L_N$), and the spin selectivities ($\Delta\sigma/\sigma_F$ and $\Delta\Sigma/\Sigma$). How can these parameters be measured in non-destructive experiments? The dc resistances $R$ discussed above cannot solve the problem. Independent experimental data like optics\cite{OO,opt} and spin e.m.f.\cite{emf} have already brought a lot of important information, and I expect the ac electrical data may also became a useful tool. However, including these phenomena into the theory involves some changes in the techniques.

All results discussed above were derived using equations for $\zeta$'a and $j$'s only, and the electrical potential $\varphi(x)$ was not involved. Therefore, the problem of the screening of electrical interactions did not appear, at least explicitly. This separation of the transport and Coulomb problems is a very special property of the dc transport in two-terminal geometry. Any generalization of the problem results in involving $\varphi(x)$. E.g., electron concentration $n(x)$ is critical for optical experiments. For small deviations from the thermodynamic equilibrium the concentrations $n_{\uparrow,\downarrow}(x)$ of up- and down-spin electrons are related to the electrochemical and electrical potentials as
\begin{equation}
\zeta_{\uparrow,\downarrow}(x)=n_{\uparrow,\downarrow}(x)/e\rho_{\uparrow,\downarrow}-\varphi(x),
\label{eq10}
\end{equation}
where $\rho_{\uparrow,\downarrow}$ are the densities of states for these electrons. Therefore, the solutions are no more universal and become depending on the dimensionality that strongly influences the screening.

In a similar way, in the ac regime the continuity equations for the spin-polarized  currents $j_{\uparrow,\downarrow}(x,t)$ include the time derivatives $\partial n_{\uparrow,\downarrow}(x,t)/\partial t$ that, quite similar to Eq.~(\ref{eq10}), bring $\varphi(x)$ into the game. Moreover, for time dependent currents the equations for the differences $\zeta_\uparrow(x,t)-\zeta_\downarrow(x,t)$ and the sums $Z(x,t)$ of the electrochemical potentials do not separate any more. In what follows, the equations for the ac response to a time dependent potential proportional to $\exp(-i\omega t)$ are presented. They were derived under the assumption that the quasineutrality condition, $n_\uparrow(x,t)+n_\downarrow(x,t)\approx 0$, is fulfilled.\cite{R02} 
 
The complex impedance of a FTN-junction ${\cal Z}(\omega)$ can be found from Eq.~(\ref{eq4}) for the dc resistance $R$ by changing the diffusion lengths $L_F$ and $L_N$ to
\end{multicols}
\widetext 
\begin{equation}
L_F(\omega)=L_F/(1-i\omega\tau_s^F)^{1/2},~~
L_N(\omega)=L_N/(1-i\omega\tau_s^N)^{1/2}.
\label{eq11}
\end{equation}
\begin{multicols}{2}
\noindent
Here $\tau_s^F$ and $\tau_s^N$ are the spin relaxation times in the F and N regions, respectively. As a result, ${\cal Z}(\omega)$ acquires a reactive part having a capacitive sign. Eq.~(\ref{eq11}) shows that two characteristic frequencies, $\omega_F=(\tau_s^F)^{-1}$ and $\omega_N=(\tau_s^N)^{-1}$, should manifest themselves in $\cal Z(\omega)$. Experimental observation of these frequencies should allow measuring the spin relaxation times. The low frequency capacitance $C_{\rm diff}$ found from Eqs.~(\ref{eq4}) and (\ref{eq11}) equals
\end{multicols}
\widetext 
\begin{equation}
C_{\rm diff}=\biggl\{\tau_s^N r_N\left(r_c{{\Delta\Sigma}\over\Sigma}
+r_F{{\Delta\sigma}\over{\sigma_F}}\right)^2 
+\tau_s^Fr_F\left[r_c{{\Delta\Sigma}\over\Sigma}
-(r_c+r_N){{\Delta\sigma}\over{\sigma_F}}\right]^2
\biggr\}/2R^2r_{FN}^2.
\label{eq12}
\end{equation}
\begin{multicols}{2}
\noindent
It is controlled by the relaxation of nonequilibrium spins and, therefore, is similar to the diffusive capacitance in the theory of $p-n$-junctions. However, the existence of the resistance $r_c$ changes the dependence of $C_{\rm diff}$ on the relaxation times. The square root dependence, $C_{\rm diff}\propto\tau_s^{1/2}$, typical of $p-n$-junctions is valid for spin injection only when $r_c=0$. In the opposite limit $r_c\gg r_F,r_N$, that is of major interest for spin injection devices, it follows from Eq.~(\ref{eq12}) that $C_{\rm diff}\propto\tau^{3/2}$. Depending from the relative magnitude of $\tau_s^F$ and $\tau_s^N$, different combinations of them can appear in $C_{\rm diff}$, and a large $\tau_s^N$ typical of semiconductor heterostructures\cite{opt} can enlarge $C_{\rm diff}$ considerably. However, it is a general regularity that a large $r_c\agt r_N, r_F$ reduces $C_{\rm diff}$.

The frequency dependences of $C_{\rm diff}(\omega)$ and of the active resistance $R_{\rm n-eq}(\omega)$ are sensitive to the relative magnitude of the basic resistances $r_F$, $r_c$ and $r_N$. Therefore, these dependences are a promising tool for measuring these resistances. Frequency dependence is also a key for separating the diffusive capacitance $C_{\rm diff}(\omega)$ from the geometric capacitance $C_{\rm geom}=\varepsilon/4\pi X$ that is expected to be frequency independent under the conditions of the 3D screening, $X$ being the contact thickness.

Similar equations can be applied to the optical experiments on the recombination of holes with electrically injected spin-polarized electrons.

Eq.~(\ref{eq3}) for spin injection is applicable also to the spin-e.m.f.\cite{emf} at a spin selective contact $\Delta\varphi_{FN}=\gamma\zeta^N_\infty/2$, where $\zeta^N_\infty$ is the difference of the potentials $\zeta_{\uparrow,\downarrow}$ in the N-region far from the contact. This $\Delta\varphi_{FN}$ includes both the contact (``valve") and the Dember contributions.

\subsection*{5~~Spin non-conserving junction}              

If spin is not conserved in a FTN-junction, because of the spin dynamics or of the spin relaxation, then the generalized Eq.~(\ref{eq1}) includes a matrix $\Sigma_{\alpha\beta}$ that is nondiagonal in the spin indeces $\alpha$ and $\beta$. The element $\Sigma_{\alpha\beta}$ describes the transfer of an electron from the $\alpha$ spin state in the ferromagnet to the $\beta$ spin state in the N-conductor. In the dissipative regime and with the time inversion symmetry violated by the spontaneous magnetization in the F-region, the only restriction imposed on these coefficients is $\Sigma_{\alpha\beta}>0$. In addition to increasing the number of parameters, the problem becomes more difficult technically also because the symmetric and antisymmetric variables, $Z(x)$ and $\zeta_\uparrow(x)-\zeta_\downarrow(x)$, do not separate any more even in the dc regime. Nevertheless, the equations of the $\gamma$-technique for the spin injection coefficients at the left and right boundaries of the contact, $\gamma_F$ and $\gamma_N$, can be derived and solved. 

Spin non-conserving junctions possess a number of peculiar properties that differ them from the spin conserving junctions discussed above. E.g., $R_{\rm n-eq}$ can change sign. This behavior is absolutely incompatible with the properties of the spin conserving junctions established above. This possibility is clearly seen in a special case when $\Sigma_{\uparrow\uparrow}=\Sigma_{\downarrow\downarrow}$, $\Sigma_{\uparrow\downarrow}=\Sigma_{\downarrow\uparrow}$, and $\Delta\sigma=0$ (what does not exclude strong bulk magnetization). Then $R_{\rm n-eq}=-\Sigma_{\uparrow\downarrow}/2\Sigma_{\uparrow\uparrow}(\Sigma_{\uparrow\uparrow}+\Sigma_{\uparrow\downarrow})<0$, hence, $R_{\rm n-eq}$ is negative. This result shows that the magnitude and even the sign of $R_{\rm n-eq}$ is controlled by a delicate balance of the processes in the bulk and at the interfaces.

\begin{center}
{\bf Acknowledgments}
\end{center}
Support from DARPA/SPINS by the Office of Naval Research Grant N000140010819 is gratefully acknowledged.

\end{multicols}
\end{document}